\documentclass[aps,prb,twocolumn]{revtex4-2}
\usepackage{amsmath,amssymb,amsfonts}
\usepackage{xcolor}
\usepackage{hyperref}
\usepackage{graphicx}
\usepackage{epsfig}
\usepackage{epstopdf}
\usepackage{color}
\usepackage{scalefnt}
\usepackage{dcolumn}
\usepackage{bm}
\usepackage{booktabs}

\begin{document}

\title{Sub-Landau-Level Structure and Correlated Electron Pairs in Quantum Hall Systems}

\author{G.-Q. Hai}
\email{hai@ifsc.usp.br}

\affiliation{Instituto de F\'{\i}sica de S\~{a}o Carlos, Universidade de S\~{a}o Paulo, 13560-970, S\~{a}o Carlos, SP, Brazil}

\begin{abstract}
We present a microscopic analysis of correlated electron motion in a two-dimensional system under a perpendicular magnetic field, starting from the interacting two-electron problem. The center-of-mass and relative motions separate exactly, and the Coulomb interaction resolves the relative-motion Landau manifold into branches labeled by the relative angular momentum $m$. We refer to these interaction-resolved branches as sub-Landau levels. Each branch retains the full center-of-mass Landau degeneracy, while its internal radial structure is continuously modified by the Coulomb interaction.
Within the lowest-Landau-level projection, the interaction contribution to each channel energy is described by the Haldane pseudopotential.
The full solution additionally contains interaction-induced radial relaxation and Landau-level admixture. We characterize these states as correlated rotating electron-pair (CREP) channels and test their relevance in fully spin-polarized $N_e=3$--6 systems with Landau-level mixing. Selected pair channels retain radial structures with fidelities $0.986$--$0.995$ to the isolated interacting two-electron solutions, and the same radial information provides useful variational ingredients for the many-electron Landau-level-mixing response. We finally discuss a possible collective channel-capacity organization leading, under a single-channel saturation assumption, to $\nu=1/|m|$. This collective counting rule is distinguished explicitly from the exact two-electron results and remains a many-body hypothesis requiring further validation.
\end{abstract}

\maketitle

\section{Introduction}

The integer and fractional quantum Hall effects (IQHE and FQHE) reveal a broad range of electronic states in two-dimensional systems under strong magnetic fields \cite{KvK,Tsui,Tong}. While the IQHE can be understood largely from single-particle Landau quantization, the FQHE is interaction driven. The Laughlin wavefunction \cite{RBL}, Haldane pseudopotentials \cite{Haldane}, hierarchy and multicomponent constructions \cite{Halperin}, and composite-fermion theory \cite{Jain} provide highly successful descriptions of the resulting many-body physics.

Despite this success, it remains valuable to explore the microscopic pictures of how Coulomb correlation
reorganizes the relative motion of electrons within the highly degenerate Landau-level manifold.
Relative angular momentum is particularly natural in this regard. After projection to a Landau level,
it labels the Haldane interaction channels, while beyond a strict projection the same quantum number
can support an interaction-dependent internal radial structure. Landau-level mixing therefore raises
a simple microscopic question: how much of the interacting two-electron radial structure remains
meaningful inside a genuine many-electron quantum Hall state?

The two-electron problem in a perpendicular magnetic field provides a controlled starting point.
Its center-of-mass and relative motions separate, and exact or quasi-exact solutions at special
interaction strengths have been studied from several viewpoints \cite{ver91,Claro,avt88,taut94,Truong00}.
Here we solve the relative problem numerically over a continuous range of the dimensionless interaction
strength $\gamma_B=\ell_B/a_B^*$ (with $\ell_B$ being the magnetic length and $a_B^*$ the effective Bohr radius)
and use the known quasi-exact solutions as benchmarks.
We call the resulting interaction- and $m$-resolved branches \emph{sub-Landau levels}.
This terminology describes the resolution of the relative-motion Landau manifold by
Coulomb interaction; it does not imply a reduction of the exact center-of-mass Landau degeneracy.

We represent the interaction-resolved radial structure through the function \(h_n^{|m|}(x)\),
with $x=r^2/2$
in the scaled relative coordinate. It reduces to the generalized Laguerre polynomial
in the noninteracting limit and describes the interaction-induced radial restructuring
at finite $\gamma_B$. In the lowest-Landau-level (LLL) projection,
the interaction contribution to the channel energy is represented by the corresponding Haldane pseudopotential.
The full two-electron solution retains the same relative-angular-momentum label while allowing radial Landau-level admixture.

We then test whether this two-electron structure survives in the many-electron problem.
Multi-Landau-level exact diagonalizations in controlled truncated bases are carried
out for fully spin-polarized systems with $N_e=3$--6. No electron-pair radial function
in the relative coordinate or permanent pair assignment is imposed on these benchmark states.
Nevertheless, selected relative-angular-momentum channels retain radial states very close
to the isolated interacting two-electron solutions, and the same radial functions provide
useful variational ingredients for describing Landau-level mixing.

Reports of correlated double-electron processes in recent relevant experiments in quantum Hall devices
provide additional  motivation for studying pair-resolved correlations \cite{Choi2015,Biswas,Demir}.
The translationally invariant model considered here does not attempt a quantitative description of those
device-specific experiments, which involve edge confinement, screening, and electrostatics
absent from the present Hamiltonian.

Finally, we distinguish the exact two-electron results and the finite-$N_e$ numerical evidence from a separate many-body proposal. A collective state dominated by a given $m$ channel may organize its available guiding-center configurations with an effective capacity proportional to $1/|m|$. A possible center-of-mass selection rule and its connection with Laughlin-type zero counting are discussed only at the many-electron level. They are not consequences of the exact two-electron separation. This separation of exact results, finite-system support, and collective hypotheses is maintained throughout the paper.

\section{Single-Particle Landau States}

For a single electron with charge $-e$ and effective mass $m_e^*$ confined to
the two-dimensional $(x,y)$ plane and subjected to a perpendicular magnetic field
${\bm B}=B\hat{\bm z}$, we use the symmetric gauge ${\bm A}={\bm B}\times{\bm r}/2$.
With the magnetic length $\ell_B=\sqrt{\hbar c/eB}$ and the cyclotron energy
$\hbar\omega_c=\hbar eB/(m_e^*c)$ as length and energy units, respectively,
the one-electron Hamiltonian is
\begin{equation}\label{H0}
\widehat H_s({\bm r})=\frac{1}{2}\left(-i{\bm\nabla}+\frac{1}{2}\hat{\bm z}\times{\bm r}\right)^2.
\end{equation}
In polar coordinates $(r,\varphi)$,
\begin{equation}\label{Esch0}
\widehat H_s({\bm r})\psi^{s}_{nm}(r,\varphi)=E^{s}_{nm}\psi^{s}_{nm}(r,\varphi),
\end{equation}
with
\begin{equation}\label{Es}
E^{s}_{nm}=E_L(n,m)=n+\frac{m+|m|+1}{2},
\end{equation}
where $n=0,1,2,\ldots$ and $m=0,\pm1,\pm2,\ldots$. The corresponding wavefunctions are
\begin{equation}\label{psis}
\psi^{s}_{nm}(r,\varphi)=\frac{e^{im\varphi}}{\sqrt{2\pi}}R^{s}_{nm}(r),
\end{equation}
with
\begin{equation}\label{Rsr}
R^{s}_{nm}(r)=\sqrt{\frac{n!}{(n+|m|)!}}
\left(\frac{r^2}{2}\right)^{|m|/2}
L_n^{|m|}\left(\frac{r^2}{2}\right)e^{-r^2/4},
\end{equation}
where $L^{|m|}_n (x)$ is the generalized Laguerre polynomial.
The orbital angular momentum is $L_z=m\hbar$. For $m\leq0$, each Landau level is macroscopically degenerate.
The number of one-electron states per unit area in a Landau level is
\begin{equation}
 n_{\phi_0}=\frac{1}{2\pi\ell_B^2}=\frac{B}{\phi_0},
\end{equation}
where $\phi_0=2\pi\hbar c/e$ is the quantum of flux, and the filling factor of the Landau levels
is $\nu=n_e/n_{\phi0}$ with $n_e$ the electron density in the 2D system.

For $N_e$ interacting electrons the Hamiltonian is
\begin{equation}\label{Hmany}
\widehat H=\sum_{i=1}^{N_e}\widehat H_s({\bm r}_i)
+\sum_{i<j}^{N_e}\frac{\gamma_B}{|{\bm r}_i-{\bm r}_j|},
\end{equation}
where
\begin{equation}
\gamma_B=\frac{\ell_B}{a_B^*},
\end{equation}
with the effective Bohr radius $a_B^*=\epsilon_r\hbar^2/m_e^*e^2$.
Equation~(\ref{Hmany}) will also be the starting point for the many-electron analysis below.
It contains all electron pairs on an equal footing and does not imply any particular pairing
of electrons. We first consider the exactly separable two-electron case.

\section{Two-Electron States and Sub-Landau-Level Structure}

\subsection{Exact separation and relative-motion eigenstates}

For two interacting electrons,
\begin{equation}\label{Hpair1}
\widehat H_p({\bm r}_1,{\bm r}_2)=\widehat H_s({\bm r}_1)+\widehat H_s({\bm r}_2)
+\frac{\gamma_B}{|{\bm r}_2-{\bm r}_1|}.
\end{equation}
We introduce the scaled center-of-mass (CM) and relative coordinates
\begin{equation}\label{cmrel}
{\bm R}=\frac{{\bm r}_1+{\bm r}_2}{\sqrt2},\qquad
{\bm r}=\frac{{\bm r}_2-{\bm r}_1}{\sqrt2}.
\end{equation}
This orthogonal scaling is chosen so that the magnetic kinetic terms of the CM and relative
Hamiltonians have the same canonical Landau form. The physical inter-electron separation is
$d\equiv|{\bm r}_2-{\bm r}_1|=\sqrt2\,r$.
The Hamiltonian separates exactly,
\begin{equation}\label{H2e}
\widehat H_p=\widehat H_{\rm cm}({\bm R})+\widehat H_{\rm rel}({\bm r}),
\end{equation}
where
\begin{equation}\label{Hcm}
\widehat H_{\rm cm}({\bm R})=\frac{1}{2}\left(-i{\bm\nabla}_{\bm R}+\frac{1}{2}\hat{\bm z}\times{\bm R}\right)^2
\end{equation}
and
\begin{equation}\label{Hrel}
\widehat H_{\rm rel}({\bm r})=\frac{1}{2}\left(-i{\bm\nabla}_{\bm r}+\frac{1}{2}\hat{\bm z}\times{\bm r}\right)^2
+\frac{\gamma_B}{\sqrt2\,r}.
\end{equation}
Thus the Coulomb term may equivalently be written as $\gamma_B/d$.

The relative Hamiltonian is rotationally invariant and $L_z^{\rm rel}$ is conserved. Its eigenstates satisfy
\begin{equation}\label{SHr}
\widehat H_{\rm rel}\psi^{\rm rel}_{nm}(r,\theta)
=E^{\rm rel}_{nm}\psi^{\rm rel}_{nm}(r,\theta),
\end{equation}
with
\begin{equation}\label{psi2e}
\psi^{\rm rel}_{nm}(r,\theta)=\frac{e^{im\theta}}{\sqrt{2\pi}}R^{\rm rel}_{nm}(r),
\qquad L_z^{\rm rel}=m\hbar.
\end{equation}
We write the radial function as
\begin{equation}\label{Rrel}
R^{\rm rel}_{nm}(r)=\sqrt{\frac{n!}{(n+|m|)!}}
\left(\frac{r^2}{2}\right)^{|m|/2}
h_n^{|m|}\!\left(\frac{r^2}{2}\right)e^{-r^2/4}.
\end{equation}
The function $h_n^{|m|}(x)$ is defined in the dimensionless Laguerre variable
$ x\equiv {r^2}/{2}={d^2}/{4} $
by the expansion
\begin{equation}\label{fhnm}
h_n^{|m|}(x)=\sum_{n'}a^{(m)}_{nn'}
\sqrt{\frac{n'!(n+|m|)!}{n!(n'+|m|)!}}\,
L_{n'}^{|m|}(x).
\end{equation}
The coefficients are obtained, for each $m$, from
\begin{equation}\label{SHrl}
\sum_{n'}\left[\left(E_L(n,m)-E^{\rm rel}_{nm}\right)\delta_{nn'}
+\gamma_B M^{|m|}_{nn'}\right]a^{(m)}_{nn'}=0,
\end{equation}
where the Coulomb matrix element in the scaled relative coordinate is
\begin{equation}\label{Mnn}
M^{|m|}_{nn'}=\frac{1}{\sqrt2}\int_0^\infty dr\,
R^s_{n'm}(r)R^s_{nm}(r).
\end{equation}
At $\gamma_B=0$, $h_n^{|m|}(x)=L_n^{|m|}(x)$ and the ordinary relative-motion Landau functions are recovered.

For special values of $\gamma_B$, the numerical solutions connect to the known quasi-exact solutions \cite{ver91,avt88,taut94,Truong00}. At such points, $h_n^{|m|}(r^2/2)$ becomes a finite polynomial in $r$. For example, $h_0^{|m|}$ is constant at $\gamma_B=0$, while at the first nontrivial quasi-exact point $\gamma_B=\sqrt{(2|m|+1)/2}$ it is linear in $r$.

Since the total two-electron wavefunction must be antisymmetric, exchange corresponds to $(r,\theta)\rightarrow(r,\theta+\pi)$. Even $m$ therefore accompanies a spin singlet, whereas odd $m$ accompanies a spin triplet. In the fully spin-polarized sector emphasized below, the LLL-like relative channels are $m=-1,-3,-5,\ldots$.

\begin{figure}[b!]
      {\includegraphics[width=8.5cm,height=7.8cm]{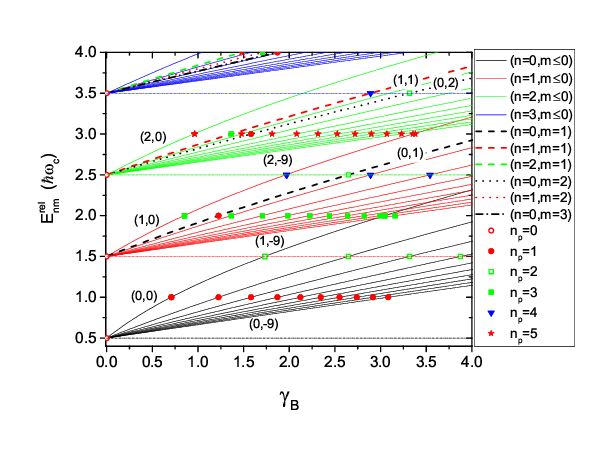}}
\vspace{-0.7cm}
       \caption{Relative-motion eigenenergy $E_{nm}^{\rm rel}$ as a function of $\gamma_B$ labeled by ($n,m$).
       The curves show the numerical solutions for the quantum numbers $n$ and $m$:
       $n=0$ (the black curves), $n=1$ (the red), $n=2$ (the green), $n=3$ (the blue),
       and $m\leq 0$ ($m=0,-1,-2,...,-9$) (the solid curves), $m$=+1 (the dashed), $m$=+2 (the dotted),
       and $m$=+3 (the dash-dotted).
       The symbols mark quasi-exact solutions indicated by $n_p$.}
       \label{Erelgam}
\end{figure}

Fig.~\ref{Erelgam} shows the relative-motion spectrum. At $\gamma_B=0$,
\begin{equation}
E^{\rm rel}_{nm}=n+\frac{m+|m|+1}{2}.
\end{equation}
For $\gamma_B>0$, the Coulomb interaction lifts the degeneracy between different
$m$ sectors while preserving $m$ as an exact quantum number. At large negative $m$
the interaction becomes weak and, to leading order,
\begin{equation}
E^{\rm rel}_{nm}\simeq n+\frac12+\frac{\gamma_B}{2\sqrt{|m|}},
\end{equation}
so $E^{\rm rel}_{nm}\rightarrow n+1/2$ as $|m|\rightarrow\infty$.

\subsection{Interaction-resolved radial structure and LLL pseudopotentials}

The relative-motion energies are directly connected with Haldane pseudopotentials \cite{Haldane}.
For the lowest radial branch, the LLL-projected interaction energy is
\begin{equation}\label{Vhaldane}
V_{|m|}=\gamma_B M^{|m|}_{00},
\end{equation}
so that in a strict LLL projection
\begin{equation}\label{ELLproj}
E^{\rm rel,LLL}_{0m}=\frac12+V_{|m|}.
\end{equation}
The full solution, by contrast, allows higher radial Landau components within the same $m$ sector.
It therefore contains information not retained in the single projected number $V_{|m|}$:
the interaction-dependent radial function $h_0^{|m|}(x)$ itself.

A useful measure of this radial relaxation is the energy lowering relative to the projected channel,
\begin{equation}\label{dErad}
\delta E_m^{\rm rad}=E^{\rm rel}_{0m}-\left(\frac12+V_{|m|}\right)\leq0.
\end{equation}
The inequality follows from the variational principle, because the full relative Hilbert space
contains the LLL-projected state as a restricted trial state.
Eq.~(\ref{dErad}) is used here only as an internal measure of radial correlation;
it is not a molecular binding energy.

The factor $r^{|m|}$ in Eq.~(\ref{Rrel}) fixes the order of the coincidence zero
and is determined by angular momentum and exchange symmetry. The additional factor
$h_0^{|m|}(x)$ describes the dynamical Coulomb restructuring of the radial state.
This distinction between kinematic zero structure and interaction-induced radial
relaxation will be important below.

\subsection{Sub-Landau levels and center-of-mass degeneracy}

The CM motion obeys
\begin{equation}\label{SHcm}
\widehat H_{\rm cm}\psi^{\rm cm}_{NM}(R,\Theta)
=E^{\rm cm}_{NM}\psi^{\rm cm}_{NM}(R,\Theta),
\end{equation}
with
\begin{equation}
E^{\rm cm}_{NM}=E_L(N,M)=N+\frac{M+|M|+1}{2}.
\end{equation}
The exact two-electron eigenstate therefore factorizes as
\begin{equation}\label{pairfactor}
\Psi_{NM,nm}({\bm R},{\bm r})
=\psi^{\rm cm}_{NM}(R,\Theta)\psi^{\rm rel}_{nm}(r,\theta),
\end{equation}
with total energy
\begin{equation}\label{Epair}
E^{\rm pair}_{NM,nm}=E^{\rm cm}_{NM}+E^{\rm rel}_{nm}.
\end{equation}
Explicitly,
\begin{align}\label{eq20}
&\Psi_{NM,nm}({\bm R},{\bm r})= \frac{1}{2\pi}
\sqrt{\frac{N!n!}{(N+|M|)!(n+|m|)!}}\; e^{iM\Theta+im\theta} \nonumber\\
&\times
\left(\frac{R^2}{2}\right)^{|M|\over 2} L^{|M|}_N\left(\frac{_{R^2}}{^2}\right)
\left(\frac{r^2}{2}\right)^{\frac{|m|}{2}} h^{|m|}_{n}\left(\frac{_{r^2}}{^2}\right)\, e^{-\frac{R^2+r^2}{4}}.
\end{align}
We refer to the interaction- and relative-angular-momentum-resolved branches
$E^{\rm pair}_{NM,nm}$ as \emph{sub-Landau levels}. The terminology emphasizes that
Coulomb interaction resolves the relative-motion Landau manifold into distinct $m$ channels.

For fixed $N,n,m$, however, the energy remains independent of every allowed negative $M$.
Each sub-Landau-level branch therefore retains the full CM Landau degeneracy. Moreover,
\begin{equation}
[H_{\rm cm},H_{\rm rel}]=[H_{\rm cm},L_z^{\rm cm}]
=[H_{\rm rel},L_z^{\rm rel}]=0,
\end{equation}
so $M$ and $m$ are independent quantum numbers of the exact two-electron problem. No relation between them, and no reduction of the exact CM degeneracy, follows from the separation in Eq.~(\ref{H2e}).

\begin{figure}[bht!]
\includegraphics[width=9.0cm,height=8.0cm]{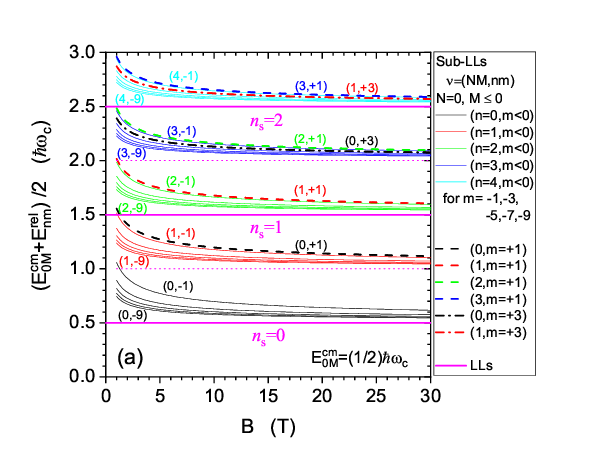}
\caption{Exact two-electron sub-Landau-level branches $E^{\rm pair}_{NM,nm}$ as a function of $B$
for $N=0$, shown as energy per electron for comparison with the single-electron Landau levels (LLs).
Only representative odd-$m$ triplet channels ($|m|\leq 9$) are displayed with $n=0$ (the black curves),
$n=1$ (the red curves), etc. The material parameters are representative of GaAs/AlGaAs
heterostructures, with \(m_e^*=0.067m_e\) and \(\epsilon_r=12.80\).}
\label{fig:subll}
\end{figure}

For the numerical illustration in Fig.~\ref{fig:subll}, we consider material parameters representative of GaAs/AlGaAs heterostructures, with an electron effective mass \(m_e^*=0.067m_e\) and a relative dielectric constant \(\epsilon_r=12.80\). These parameters correspond to an effective Bohr radius \(a_B^*\simeq10.11\) nm and an interaction parameter \(\gamma_B\simeq2.54/\sqrt{B[\mathrm{T}]}\). The sub-Landau-level energies are shown in units of \(\hbar\omega_c\), with the pair energies divided by two for comparison with the single-electron Landau levels.
Fig.~\ref{fig:subll} illustrates the exact sub-Landau-level branches. For increasingly negative $m$, the Coulomb contribution decreases and the branches approach the corresponding noninteracting Landau energy quasi-continuously. The exact two-electron solution therefore establishes two distinct structures: the conventional CM guiding-center degeneracy and an interaction-dependent organization of the relative motion into $m$-resolved correlation channels.

\section{Correlated Rotating Electron Pairs and Relative Angular Momentum}

\subsection{CREP as an interaction-resolved relative-motion state}

The exact solutions of Sec.~III provide a microscopic description of correlation within a fixed relative-angular-momentum channel. For $m<0$, the two electrons have a definite relative orbital angular momentum $L_z^{\rm rel}=m\hbar$, and their relative probability density is distributed around the common center of mass. We refer to such an interacting relative-motion eigenstate as a \emph{correlated rotating electron pair} (CREP).

The term CREP characterizes the internal two-electron correlation structure of an exact relative-motion eigenstate. It does not imply a permanently assigned molecular pair in a many-electron system, nor does it assume that interactions with the remaining electrons are weak. For the lowest radial branch,
\begin{equation}\label{crepwave}
\psi^{\rm rel}_{0m}(r,\theta)\propto e^{im\theta}r^{|m|}
h_0^{|m|}\!\left(\frac{r^2}{2};\gamma_B\right)e^{-r^2/4}.
\end{equation}
The factor $r^{|m|}$ contains the symmetry-determined coincidence zero, whereas $h_0^{|m|}$ contains the additional interaction-induced radial restructuring. We use \emph{dynamical Coulomb correlation} for the latter effect, distinguishing it from exchange consequences imposed by fermionic antisymmetry.

\subsection{Physical picture of the relative motion}

\begin{figure}[t]
\includegraphics[width=7.0cm,height=4.0cm]{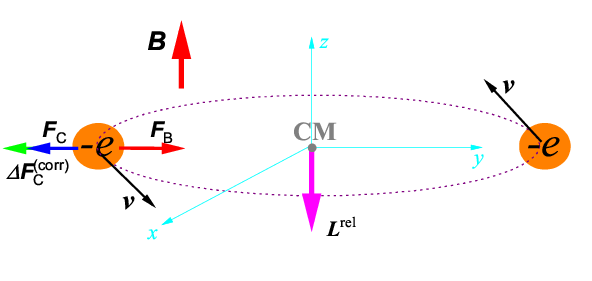}
\caption{Schematic semiclassical representation of a CREP with ${\bm L}^{\rm rel}=m\hbar\hat{\bm z}$ ($m<0$).
The arrows ${\bm F}_{\rm C}$ and ${\bm F}_{\rm B}$ denote the Coulomb interaction and magnetic contribution
to the intuitive force picture, while the arrow \(\Delta {\bm F}_{\rm C}^{(\rm{corr})}\) schematically
represents a correlation-induced modification of the effective Coulomb force relative to a reference state,
without implying an additional fundamental interaction.}
\label{fig:crep}
\end{figure}

A useful intuitive representation of a CREP is shown in Fig.~\ref{fig:crep}. The two electrons move
in a perpendicular magnetic field, interact through the Coulomb interaction, and carry
a definite relative angular momentum. In a semiclassical picture, the magnetic field sets the sense
of orbital circulation while Coulomb repulsion and the correlated spatial distribution
determine the internal structure of the state.
Notice that electron correlation does not produce an additional fundamental interaction beyond the Coulomb
term in Eq.~(\ref{Hpair1}). Correlation holes and correlation-induced redistribution of
electronic density are standard ways of describing this effect in atomic and molecular
systems \cite{Coulson,hattig,Hai23,Hai26}. In the present magnetic two-electron problem,
the same physical idea is encoded directly in the interaction dependence of $h_0^{|m|}(x)$.
Nevertheless, the effective force due to correlation symbolized
as the arrow  \(\Delta {\bm F}_{\rm C}^{(\rm{corr})}\) schematically represents a
correlation-induced modification of the effective Coulomb force relative to a reference state.
Its direction and magnitude are not quantitatively determined here.

However, it is important to remember that
the angular factor $e^{im\theta}$ carries an integer phase winding around the coincidence point $r=0$.
Thus a fixed-$m$ relative channel also carries a well-defined pair-vortex winding number $m$.

\subsection{Correlation energy and scope of the CREP interpretation}

The energetic effect of radial dynamical correlation can be quantified internally through Eq.~(\ref{dErad}), without importing a correlation-energy scale from atomic or molecular systems. The quantity $\delta E_m^{\rm rad}$ measures the lowering obtained when higher radial Landau components are allowed within a fixed $m$ channel. It should not be interpreted as a molecular binding energy or as defining a pair-dissociation temperature.

In a {\it many-electron environment}, a CREP is therefore best regarded as a correlation-resolved pair channel rather than as a permanently bound object. Whether a particular $m$ channel acquires large weight is a collective property of the many-electron state and must be determined independently. This point motivates the finite-system tests in Sec.~VI.

\section{Zeeman Splitting, Spin Polarization, and Disorder}

The orbital two-electron problem fixes the connection between spin symmetry and relative angular momentum: even $m$ accompanies a spin singlet and odd $m$ a spin triplet. The Zeeman interaction further splits the spin sectors. For two electrons,
\begin{equation}\label{EZ}
E_Z=2\Delta_Z S_z,
\end{equation}
where
\begin{equation}
\Delta_Z=\frac{g^*\mu_BB}{2}=g^*\frac{m_e^*}{4m_e}\hbar\omega_c.
\end{equation}
For a singlet $S_z=0$, whereas a triplet has $S_z=0,\pm1$. For GaAs, using $m_e^*=0.067m_e$ and $g^*\simeq-0.44$, one obtains $\Delta_Z\simeq-0.00737\hbar\omega_c$. The $S_z=+1$ triplet component is therefore lowered relative to the $S_z=0$ and $S_z=-1$ triplet components.

The competition between singlet and triplet states is determined by the total energy, including their different orbital correlation energies and the Zeeman contribution. Zeeman splitting alone does not destabilize a singlet, whose total Zeeman energy remains zero. We therefore do not infer a universal singlet--triplet transition from the Zeeman scale alone. In the fully spin-polarized regime relevant to the principal odd-denominator states, the pair channels emphasized below are
\begin{equation}
m=-1,-3,-5,\ldots.
\end{equation}

Disorder broadens Landau levels and mixes guiding-center states. If the broadening becomes comparable to the energy differences between competing correlation channels, their spectral distinction can become difficult to resolve. Cleaner samples therefore provide a more favorable setting in which interaction-resolved channel structure can influence the observed quantum Hall state. We do not, however, identify a universal disorder threshold for the existence or destruction of a CREP, because the CREP is defined here as a pair-correlation channel rather than as a bound molecular object.

\section{Many-Electron Extension and Finite-System Validation}

The many-electron starting point remains the full Hamiltonian in
Eq.~(\ref{Hmany}), where every electron interacts with every other
one. No permanent assignment into disjoint pairs is implied.

The exact two-electron solutions nevertheless provide a natural set of
correlation-resolved pair channels labeled by $m$. In a many-electron
state these channels are components of the two-particle reduced
structure, not distinct molecular pairs. For any pair $(i,j)$ one may
project the relative motion onto fixed-$m$ sectors and examine both
the weight of the channel and its internal radial state. The central
question is therefore not whether the electrons form permanent CREPs,
but whether the interacting two-electron radial structure remains
identifiable inside the full many-electron wavefunction.

\subsection{Finite-$N_e$ validation of the two-electron radial structure}

To test this question, we performed multi-Landau-level exact
diagonalizations for fully spin-polarized systems with
\begin{equation}
N_e=3,4,5,6,
\end{equation}
including the kinetic energy and bare Coulomb interaction. The
calculations are carried out in fixed-total-angular-momentum sectors
\begin{equation}\label{Jsector}
L_z=-J\hbar,\qquad J=\frac{3N_e(N_e-1)}{2},
\end{equation}
so that $J=9,18,30,45$ for $N_e=3,4,5,6$, respectively. These are the
angular-momentum sectors of the corresponding $\nu=1/3$ Laughlin
droplets; their selection does not assume that the multi-Landau-level
eigenstates have Laughlin form. The basis contains all Slater
configurations in the chosen sector satisfying
\begin{equation}\label{Kcut}
\sum_i a_i\leq K,
\end{equation}
where $a_i$ is the one-electron cyclotron occupation. Thus $K$ is a
cutoff on the total cyclotron excitation, not a fixed number of complete
single-particle Landau levels. No confinement, finite thickness,
disorder, CREP radial function, or permanent pair assignment is imposed
on these benchmark states.

We normalize the two-particle reduced density matrix to unit trace,
$\operatorname{Tr}\rho^{(2)}=1$, with the average taken over all
unordered electron pairs. After tracing over the spectator and pair
center-of-mass degrees of freedom, let $\rho_m$ denote its radial block
in the relative-angular-momentum sector $m$. The channel weights
therefore satisfy $\sum_m w_m=1$, with the sum over all allowed
channels. We define
\begin{equation}\label{Fmdef}
w_m=\operatorname{Tr}\rho_m,\qquad
R_m=\frac{\rho_m}{w_m},\qquad
F_m={\bm c}_m^\dagger R_m{\bm c}_m,
\end{equation}
where ${\bm c}_m$ is the normalized isolated interacting two-electron
radial ground state at the same $\gamma_B$ and $m$. Thus $w_m$ is the
probability weight of the selected channel and $R_m$ has unit trace for
$w_m>0$. The fidelity $F_m$ measures radial agreement conditional on
that channel; it is not a full many-electron overlap.

\begin{table}[t]
\caption{Conditional pair-channel weights and radial fidelities at $\gamma_B=3.5$.
The basis cutoff $K$ is defined by Eq.~(\ref{Kcut}).}
\label{tab:pairfidelity}
\centering
\small
\setlength{\tabcolsep}{3.3pt}
\begin{tabular}{cccccc}
\toprule
$N_e$ & $K$ &\;\;  $w_{-3}$ & $F_{-3}$ &\;\;  $w_{-5}$ & $F_{-5}$\\
\midrule
3 & 12 &\;\;  0.38791 & 0.99136 &\;\;  0.43183 & 0.99417\\
4 & 8  &\;\;  0.28695 & 0.99261 &\;\;  0.28488 & 0.98826\\
5 & 5  &\;\;  0.27140 & 0.99512 &\;\;  0.19680 & 0.98605\\
6 & 4  &\;\;  0.21363 & 0.99338 &\;\;  0.20616 & 0.99119\\
\bottomrule
\end{tabular}
\end{table}

As shown in Table~\ref{tab:pairfidelity}, at $\gamma_B=3.5$ the
conditional fidelities for the $m=-3$ and $m=-5$ channels lie in the
range
\begin{equation}
0.9860\lesssim F_m\lesssim0.9951
\end{equation}
for all $N_e=3$--6 systems studied. For $N_e=5$ and 6, the
corresponding fidelities with the unmixed LLL radial states at the same
coupling are only $0.9295$--$0.9535$. The closer agreement is therefore
specifically associated with the interaction-induced radial
restructuring of the exact two-electron states, not merely with the
common angular-momentum label.

The calculations use computationally accessible cutoffs and are
monitored by successive-$K$ convergence. For the most demanding case,
$N_e=6$ and $\gamma_B=3.5$, the squared overlap between the $K=3$ and
$K=4$ benchmark states is $0.999378$, while the final energy change is
$2.24\times10^{-3}\hbar\omega_c$. These are finite-basis convergence
diagnostics rather than rigorous bounds on the remaining truncation
error.

The results show that selected $m$ channels retain an internal radial
state extremely close to the isolated interacting two-electron solution
even when all electron pairs interact simultaneously. They support the
use of CREP states as correlation-resolved pair-channel building
blocks, but they do not imply a decomposition of the many-electron
ground state into permanent pairs.

\subsection{Variational usefulness for Landau-level mixing}

We next examine whether the same two-electron radial information can be
used constructively. The normalized reference state $\Phi_0$ is the
Coulomb ground state within the LLL in the chosen $J$ sector; it is
already correlated and is not assumed to be the analytic Laughlin
state. To describe interaction-induced radial excitations, let
$c_{qm}\equiv a_{0q}^{(m)}$ denote the coefficients of the isolated
interacting two-electron radial ground state expanded in noninteracting
relative states $|q,m\rangle$, as in Eq.~(\ref{fhnm}). The reference
amplitude is $c_{0m}$. For each odd $m<0$ we introduce
\begin{equation*}
\begin{aligned}
\widehat Q_m={}P_{J,K}\sum_{i<j}\sum_{q\geq1}
\frac{c_{qm}}{c_{0m}}
\bigl(|q,m\rangle\langle0,m|\bigr)_{ij}P_{J,K},
\end{aligned}
\end{equation*}
where $P_{J,K}$ projects onto the fixed-$J$, total-excitation-cutoff
basis of Eqs.~(\ref{Jsector}) and (\ref{Kcut}). The pair operator acts
on the relative coordinate of electrons $i,j$, as the identity on
their center-of-mass and spectator coordinates. Only transitions
allowed by the cutoff are retained. The compact variational space is
\begin{equation*}
\begin{aligned}
\mathcal V_{J,K}=\operatorname{span}\{\Phi_0,
\widehat Q_{-1}\Phi_0,\widehat Q_{-3}\Phi_0,
\widehat Q_{-5}\Phi_0,
\widehat Q_{\leq-7}\Phi_0\},
\end{aligned}
\end{equation*}
with $\widehat Q_{\leq-7}=\sum_{m=-7,-9,\ldots}\widehat Q_m$ over the
channels supported in the sector. The operators act on all electron
pairs, so no permanent pair covering is introduced. The trial state
$\Psi_{\rm trial}$ is the lowest-energy Rayleigh--Ritz vector obtained
from the Hamiltonian and overlap matrices in this generally
nonorthogonal five-vector space. The same physical Hamiltonian and
cutoff are used for the trial and benchmark calculations. No
variational parameter is fitted to the multi-Landau-level benchmark
energy or wavefunction; the benchmark is used only for comparison.

We characterize the performance by
\begin{equation}\label{etadef}
\eta=\frac{E_{\rm LLL}-E_{\rm trial}}{E_{\rm LLL}-E_{\rm full}},
\qquad
F_{\rm MB}=|\langle\Psi_{\rm trial}|\Psi_{\rm full}\rangle|^2.
\end{equation}
Here $E_{\rm trial}$ is the minimum Rayleigh--Ritz energy,
$\eta$ is the fraction of the benchmark Landau-level-mixing energy
lowering recovered by the variational state, and $F_{\rm MB}$ is its
squared many-electron overlap with the corresponding multi-Landau-level
benchmark.

At $\gamma_B=1$, the compact trial space recovers about $96$--$97\%$
of the benchmark Landau-level-mixing lowering for $N_e=5$ and 6, with
many-body fidelities close to unity. At $\gamma_B=3.5$, where Landau-level
mixing is stronger, the recovery remains approximately $76$--$81\%$,
although the reduced full-state overlaps show that a richer
many-electron variational space is then required. The high conditional
fidelities in Table~\ref{tab:pairfidelity} and the lower $F_{\rm MB}$
values at stronger coupling are not contradictory: the former test
the radial structure inside a selected pair channel, whereas the
latter test the complete many-electron wavefunction.

These results show that the isolated interacting radial functions are
not merely diagnostics; they capture an appreciable part of the actual
many-electron Landau-level-mixing response. A systematic construction
of more complete effective many-body interactions from this
observation is beyond the scope of the present work.

\begin{table}[tbh]
\caption{Variational performance of the compact pair-channel radial
construction for $N_e=5,6$. Here $K$ is the total cyclotron-excitation
cutoff, $\eta$ is the fraction of Landau-level-mixing energy lowering
recovered, and $F_{\rm MB}$ is the squared many-electron fidelity with
the corresponding multi-Landau-level benchmark.}
\label{tab:variational}
\centering
\small
\setlength{\tabcolsep}{5pt}
\begin{tabular}{ccccc}
\toprule
$N_e$ & $\gamma_B$ & $K$ &\hspace{0.5cm} $\eta$ & $F_{\rm MB}$\\
\midrule
5 & 1.0 & 5 &\hspace{0.5cm} 96.961\% & 0.997289\\
6 & 1.0 & 3 &\hspace{0.5cm} 95.974\% & 0.996023\\
5 & 3.5 & 5 &\hspace{0.5cm} 80.798\% & 0.921526\\
6 & 3.5 & 4 &\hspace{0.5cm} 75.819\% & 0.871248\\
\bottomrule
\end{tabular}
\end{table}

\subsection{Collective channel organization and effective capacity}
\label{sec:capacity}

The exact two-electron problem establishes that $M$ and $m$ are
independent. In a many-electron system, however, antisymmetry,
collective occupation, and spatial exclusion associated with a
correlation hole constrain how pair channels can be accommodated. This
motivates asking whether collective states can favor particular
relations between pair center-of-mass guiding-center labels and
relative-angular-momentum channels.

\begin{figure}[b]
\centering
\includegraphics[width=0.98\columnwidth]{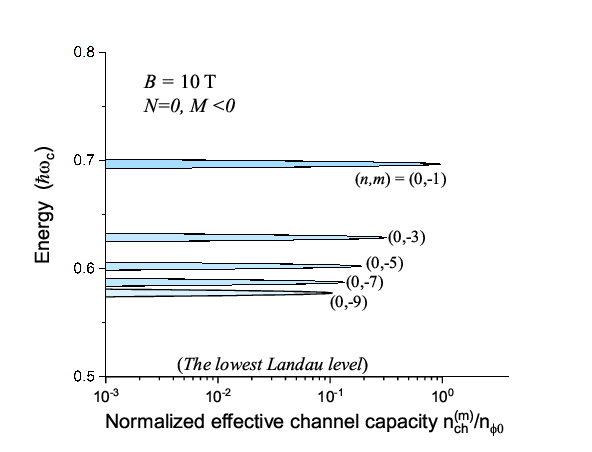}
\caption{Schematic effective channel-capacity spectrum for
$m=-1,-3,\ldots,-9$, with $n_{\rm ch}^{(m)}=n_{\phi0}/|m|$. The vertical
positions represent the corresponding channel energies at $B=10$~T
shown in Fig.~2. The finite vertical widths are phenomenological and
illustrate possible many-electron broadening due to channel mixing,
Landau-level mixing, disorder, and related collective effects. This
panel does not represent an exact two-electron density of states.}
\label{fig:channelcapacity}
\end{figure}

One possible organization is the sequence
\begin{equation}\label{Mmproposal}
M=M_{\rm ref}+km,\qquad k=0,1,2,\ldots,
\end{equation}
where $M_{\rm ref}$ specifies a reference CM sector. Unlike the
isolated two-electron case, the CM angular momentum of an individual
pair is not generally a conserved quantum number of the full
many-electron Hamiltonian. Equation~(\ref{Mmproposal}) must therefore
be interpreted as a proposed collective organization of pair-channel
occupations, not as a simultaneous exact quantum-number relation for
every pair. The proposed selection pattern is not derived from the
exact two-electron Hamiltonian, and its validity requires independent
many-electron evidence.

Correspondingly, we define a proposed effective \emph{channel capacity
per unit area}
\begin{equation}\label{nch}
n_{\rm ch}^{(m)}=\frac{n_{\phi0}}{|m|}
=\frac{1}{2\pi|m|\ell_B^2}.
\end{equation}
The quantity $n_{\rm ch}^{(m)}$ is not a reduced exact degeneracy of a
two-electron sub-Landau level. Rather, it parametrizes the number of
effectively available electron guiding-center states in a many-electron
fluid \emph{assumed} to be organized predominantly through a single
$m$ channel. This conditional interpretation is illustrated in
Fig.~\ref{fig:channelcapacity}. The figure shows the capacity at the
corresponding channel energies; its finite widths are schematic and do
not indicate a lifting of exact two-electron CM degeneracy. They
illustrate possible many-body broadening from channel hybridization,
Landau-level mixing, disorder, and related collective effects.

If a single channel is saturated according to Eq.~(\ref{nch}), its
electron density is $n_e=n_{\rm ch}^{(m)}$ and the corresponding filling
is
\begin{equation}\label{nucapacity}
\nu=\frac{n_e}{n_{\phi0}}=\frac{1}{|m|}.
\end{equation}
Thus $m=-1,-3,-5,\ldots$ are associated, within this proposed
single-channel capacity construction, with $\nu=1,1/3,1/5,\ldots$.
The result is a many-electron channel-counting hypothesis, not a new
two-electron degeneracy statement or an established microscopic
occupation distribution of a Laughlin state.

\subsection{Relation to Laughlin-type states and limitations}

It is useful to make the connection with Laughlin-type zero structure
explicit. We adopt the complex coordinate
\begin{equation}
z=x+iy,\qquad \bar z=x-iy.
\end{equation}
With the present convention $m<0$ for the LLL-like states, the
two-electron short-distance factor is naturally written in terms of
$\bar z$,
\begin{equation}\label{pairzero}
\psi_m(z_i,z_j)\sim(\bar z_i-\bar z_j)^{|m|},
\qquad z_i\rightarrow z_j.
\end{equation}
The order-$|m|$ coincidence zero is fixed by relative angular momentum,
whereas the radial factor $h_0^{|m|}(d_{ij}^2/4;\gamma_B)$ carries the
additional Coulomb-induced restructuring.

If a uniform many-electron state realizes the same minimum order-$|m|$
coincidence zero for every electron pair, a natural interaction-dressed
trial structure is
\begin{align}\label{manycrep}
\Psi_{\rm CREP}^{(m)}(\{z_i\})\propto&
\prod_{i<j}(\bar z_i-\bar z_j)^{|m|}
\prod_{i<j}h_0^{|m|}\!\left(\frac{d_{ij}^2}{4};\gamma_B\right)\nonumber\\
&\times\exp\left[-\frac14\sum_i|z_i|^2\right].
\end{align}
The variational calculations of Sec.~VI.B test a distinct
radial-dressing construction based on the correlated
LLL reference state $\Phi_0$; they should not be
interpreted as a direct numerical validation of
the illustrative all-pair trial form in the above equation.
Equation~(\ref{manycrep}) is suggested by the two-electron structure;
it is not derived as the unique many-electron ground state. When the
radial factors are replaced by constants, it reduces to the Laughlin
polynomial structure at $\nu=1/|m|$ \cite{RBL}. In this sense the
Laughlin state realizes a uniform minimum-order pairwise-zero
structure in the LLL, while the CREP radial factors suggest a way to
incorporate interaction-induced radial relaxation beyond strict LLL
projection.

This connection is structural rather than a derivation of Laughlin
order from the isolated two-electron problem. The two-electron solution
does not require every pair in a many-electron state to occupy the same
channel. In general one expects a distribution of pair-channel weights
$P_m$, with
\begin{equation}
\sum_m P_m=1.
\end{equation}
When $P_m$ is obtained from the same unit-trace two-particle density
matrix used in Sec.~VI.A, it coincides with the channel weight $w_m$
of Eq.~(\ref{Fmdef}). For a Laughlin state with odd exponent $p$,
relative-angular-momentum channels with odd $|m|<p$ are absent,
whereas channels with $|m|\geq p$ can remain populated. In particular,
a uniform order-$p$ coincidence-zero condition does \emph{not} imply
that every pair occupies only $m=-p$ (for $p=3$, $P_{-1}=0$, while
$P_{-3},P_{-5},\ldots$ can be nonzero). The single-channel capacity
construction of Sec.~VI.C is therefore a distinct many-electron
hypothesis, not an exact statement about the Laughlin channel
occupation distribution.

For identical antisymmetric geminals, full antisymmetrization of a
product state has a Pfaffian structure, but a Pfaffian alone neither
produces the Laughlin all-pair order-$|m|$ zeros nor establishes
Moore--Read topological order \cite{MooreRead}. Likewise, Halperin
multicomponent states contain collective correlations and internal
degrees of freedom that are not fixed by an isolated two-electron
radial solution \cite{Halperin}.

The status of the results can therefore be summarized at three
distinct levels. The CM-relative separation and interacting
two-electron radial states are exact within the two-electron problem.
Their persistence and variational usefulness for $N_e=3$--6 are
supported numerically within controlled truncated multi-Landau-level
bases. The collective $M$--$m$ selection rule, channel capacity, and
single-channel saturation are many-electron organizing hypotheses
whose thermodynamic validity remains to be established. The CREP
framework is used here as a microscopic description of
interaction-resolved pair correlations, not as a claim that a quantum
Hall liquid consists of permanently bound electron pairs.

\section{Discussion}

The results presented here suggest a microscopic organization of quantum Hall correlations in terms of interaction-resolved two-electron relative-motion channels. The starting point is exact: the CM and relative coordinates separate for two electrons in a perpendicular magnetic field, Coulomb interaction acts in the relative sector, and the relative Landau manifold is resolved into $m$-dependent branches while the full CM degeneracy remains intact. The term sub-Landau level is used for these branches, not for a new reduction of the exact Landau degeneracy.

The important microscopic content of a channel is not only its energy but also its radial structure. In a strict LLL projection, a relative-angular-momentum channel is characterized energetically by the Haldane pseudopotential $V_{|m|}$ \cite{Haldane}. In the full two-electron problem, Coulomb interaction additionally produces radial Landau-level admixture while preserving $m$. The function $h_n^{|m|}(x)$ provides a compact representation of this interaction-dressed radial structure. Thus the CREP viewpoint is complementary to the pseudopotential description: it resolves the internal real-space structure associated with the same relative-angular-momentum channel.

The finite-system calculations provide the bridge to the many-electron problem that was absent from a purely two-electron argument.
For the selected total-angular-momentum sectors of fully spin-polarized \(N_e=3\)–6 systems, the \(m=-3\) and \(m=-5\) channels retain conditional radial fidelities of \(0.986\)–\(0.995\) relative to the isolated interacting two-electron solutions at \(\gamma_B=3.5\). The same radial information also supplies useful variational ingredients for Landau-level mixing. These results do not imply permanent electron pairs. They show instead that the interaction-resolved two-electron states remain meaningful components of the two-particle sector of the many-electron wavefunction.

The relation to the Laughlin description is likewise complementary. The factor $(\bar z_i-\bar z_j)^{|m|}$ contains the same pairwise coincidence-zero structure that underlies Laughlin correlations, while $h_0^{|m|}$ contains additional radial information outside the purely polynomial LLL form.
For the Laughlin state at the maximal density compatible with the uniform order-$|m|$ pairwise-zero condition,
the usual zero counting yields $\nu=1/|m|$ in the thermodynamic limit. The proposed effective channel
capacity of Eq.~(42) gives the same filling sequence under a separate collective-occupation assumption.
This agreement does not establish the proposed $M$--$m$ selection rule.

A separate open question is whether the collective occupation dynamically selects a relation such as Eq.~(\ref{Mmproposal}). The exact two-electron Hamiltonian does not do so. We therefore regard the $M$--$m$ relation and Eq.~(\ref{nch}) as hypotheses about many-electron organization. Their usefulness must ultimately be judged by larger-system channel distributions, competing angular-momentum sectors, and thermodynamic behavior.

The present perspective also differs in scope from composite-fermion theory. Composite fermions provide a highly successful global organization of fractional quantum Hall states and their hierarchy \cite{Jain}. Relative-angular-momentum-dependent residual interactions can themselves lead to pairing tendencies among composite fermions \cite{Scarola}. The CREP construction addresses a more microscopic bare-electron question: how Coulomb interaction restructures the internal radial motion within an electron pair channel and whether that structure survives inside the many-electron state. A direct identification of a bare-electron CREP channel with a composite-fermion channel is therefore not implied. Establishing a quantitative connection, including hierarchy states such as $\nu=2/5$, is an interesting future problem.

Pair- and cluster-resolved correlations are also relevant in charge-ordered quantum Hall regimes. Bubble phases provide established examples in which several carriers occupy a common density maximum \cite{Koulakov}, and paired Wigner-crystal constructions illustrate how local two-electron structure can enter a collective crystalline state \cite{TautPWC}. The present calculations do not establish any such phase; they only suggest that the channel-resolved radial analysis could be useful when studying local correlations in them.

The present Hamiltonian assumes a parabolic band. In graphene, the Dirac spinor structure, Landau-level form factors, and spin/valley/sublattice degrees of freedom require a separate treatment \cite{Goerbig}. The radial equations derived here therefore cannot be transferred by a simple effective-mass substitution. The broader idea of resolving interaction effects through pair channels may nevertheless be useful in such multicomponent Landau-quantized systems.

Several questions remain open. The thermodynamic distribution of pair-channel weights $P_m$ has not been derived from the present finite-size calculations. The proposed $M$--$m$ selection rule and effective channel capacity require independent many-body validation. A systematic description of channel mixing and induced higher-body effects remains to be developed, as does the role of excited radial branches in higher-order fractional states. These limitations do not alter the controlled results established here: the exact interacting two-electron radial structure, its persistence in the finite many-electron systems studied, and its variational relevance for Landau-level mixing.

The main physical implication is therefore that relative angular momentum carries more information than a label for a projected interaction. Once radial Landau-level mixing is allowed, each $m$ channel acquires an interaction-dependent internal structure, and the finite-$N_e$ results show that this structure remains detectable inside correlated many-electron states. This provides a microscopic bridge between the exactly separable two-electron problem and the pair correlations that organize quantum Hall many-body states.

\section{Conclusion}

We have analyzed the interacting two-electron problem in a perpendicular magnetic field as a microscopic starting point for quantum Hall correlation. Exact CM-relative separation leads to interaction-resolved relative-angular-momentum channels whose radial structure is continuously modified by the Coulomb interaction. These branches, which we call sub-Landau levels, retain the full CM Landau degeneracy; the exact two-electron problem imposes no relation between the CM and relative angular momenta.

The relevance of this structure to the many-electron problem has been tested through multi-Landau-level exact diagonalization in controlled truncated bases for fully spin-polarized $N_e=3$--6 systems. Selected $m$ channels retain radial states with very high fidelity to the corresponding isolated interacting two-electron solutions, and the same radial information provides useful variational ingredients for the Landau-level-mixing response. These results support the interpretation of CREP states as correlation-resolved pair channels or microscopic building blocks, rather than as permanently bound electron pairs.

A separate many-body question concerns the collective occupation of these channels. We have proposed an
effective channel capacity proportional to $1/|m|$, whose saturation would yield the filling sequence
$\nu=1/|m|$. This sequence is consistent with Laughlin zero counting, although the two descriptions do not
imply identical pair-channel occupations. The proposed collective $M$--$m$ organization is not derived from the
exact two-electron Hamiltonian and remains a hypothesis requiring further many-body validation.

The present work therefore establishes a controlled bridge between exact two-electron correlation physics and finite many-electron quantum Hall states. Extending this bridge to mixed-channel and hierarchy states, and developing a systematic treatment of the induced many-body interactions, are natural directions for future work.

\vspace{-0.5cm}
\begin{acknowledgments}
This work was supported by FAPESP (S\~ao Paulo Research Foundation, under Grant No.~2024/00484-2) and CNPq (Brazil).
\end{acknowledgments}


\begin{thebibliography}{10}

\bibitem{KvK} K. von Klitzing, G. Dorda, and M. Pepper,
New Method for High-Accuracy Determination of the Fine Structure Constant Based on Quantized Hall Resistance,
Phys. Rev. Lett. \textbf{45}, 494 (1980).

\bibitem{Tsui} D. C. Tsui, H. L. Stormer, and A. C. Gossard,
Two-Dimensional Magnetotransport in the Extreme Quantum Limit,
Phys. Rev. Lett. \textbf{48}, 1559 (1982).

\bibitem{Tong} D. Tong, Lectures on the Quantum Hall Effect, arXiv:1606.06687 [hep-th] (2016).

\bibitem{RBL} R. B. Laughlin,
Anomalous Quantum Hall Effect: An Incompressible Quantum Fluid with Fractionally Charged Excitations,
Phys. Rev. Lett. \textbf{50}, 1395 (1983).

\bibitem{Haldane} F. D. M. Haldane,
Fractional Quantization of the Hall Effect: A Hierarchy of Incompressible Quantum Fluid States,
Phys. Rev. Lett. \textbf{51}, 605 (1983).

\bibitem{Halperin} B. I. Halperin,
Theory of the Quantized Hall Conductance,
Helv. Phys. Acta \textbf{56}, 75 (1983).

\bibitem{Jain} J. K. Jain,
Composite-Fermion Approach for the Fractional Quantum Hall Effect,
Phys. Rev. Lett. \textbf{63}, 199 (1989).

\bibitem{ver91} A. Ver\c{c}in,
Two Anyons in a Static, Uniform Magnetic Field. Exact Solution,
Phys. Lett. B \textbf{260}, 120 (1991).

\bibitem{Claro} S. Curilef and F. Claro,
Dynamics of Two Interacting Particles in a Magnetic Field in Two Dimensions,
Am. J. Phys. \textbf{65}, 244 (1997).

\bibitem{avt88} A. V. Turbiner,
Quasi-Exactly-Solvable Problems and $sl(2)$ Algebra,
Commun. Math. Phys. \textbf{118}, 467 (1988).

\bibitem{taut94} M. Taut,
Two Electrons in a Homogeneous Magnetic Field: Particular Analytical Solutions,
J. Phys. A: Math. Gen. \textbf{27}, 1045 (1994); \textbf{27}, 4723 (1994) (erratum).

\bibitem{Truong00} T. T. Truong and D. Bazzali,
Exact Low-Lying States of Two Interacting Equally Charged Particles in a Magnetic Field,
Phys. Lett. A \textbf{269}, 186 (2000).

\bibitem{Choi2015} H. K. Choi, I. Sivan, A. Rosenblatt, M. Heiblum, V. Umansky, and D. Mahalu,
Robust Electron Pairing in the Integer Quantum Hall Effect Regime,
Nat. Commun. \textbf{6}, 7435 (2015).

\bibitem{Biswas} S. Biswas, H. K. Kundu, V. Umansky, and M. Heiblum,
Electron Pairing of Interfering Interface-Based Edge Modes,
Phys. Rev. Lett. \textbf{131}, 096302 (2023).

\bibitem{Demir} A. Demir, N. Staley, S. Aronson, S. Tomarken, K. West, K. Baldwin, L. Pfeiffer, and R. Ashoori,
Correlated Double-Electron Additions at the Edge of a Two-Dimensional Electronic System,
Phys. Rev. Lett. \textbf{126}, 256802 (2021).

\bibitem{Coulson} C. A. Coulson and A. H. Neilson,
Electron Correlation in the Ground State of Helium,
Proc. Phys. Soc. \textbf{78}, 831 (1961).

\bibitem{hattig} C. H\"attig, W. Klopper, A. K\"ohn, and D. P. Tew,
Explicitly Correlated Electrons in Molecules,
Chem. Rev. \textbf{112}, 4 (2012).

\bibitem{Hai23} G.-Q. Hai, L. C\^andido, B. G. A. Brito, and Y. Liu,
Revealing the Regularities of Electron Correlation Energies Associated with Valence Electrons
in Atoms in the First Three Rows of the Periodic Table,
Chem. Phys. Lett. \textbf{855}, 141567 (2024).

\bibitem{Hai26}G.-Q. Hai, L. C\^andido, and B. G. A. Brito,
Electron Correlation in Chemical Bonding: From Dispersion to Covalent Stabilization in Prototype Diatomic Systems,
J. Phys. Chem. A \textbf{130}, 7875–7886 (2026).

\bibitem{MooreRead} G. Moore and N. Read,
Nonabelions in the Fractional Quantum Hall Effect,
Nucl. Phys. B \textbf{360}, 362 (1991).

\bibitem{Scarola} V. W. Scarola, K. Park, and J. K. Jain,
Cooper Instability of Composite Fermions,
Nature \textbf{406}, 863 (2000).

\bibitem{Koulakov} A. A. Koulakov, M. M. Fogler, and B. I. Shklovskii,
Charge Density Wave in Two-Dimensional Electron Liquid in Weak Magnetic Field,
Phys. Rev. Lett. \textbf{76}, 499 (1996).

\bibitem{TautPWC} M. Taut,
On the Paired Wigner Crystal,
Solid State Commun. \textbf{89}, 189 (1994).

\bibitem{Goerbig} M. O. Goerbig,
Electronic Properties of Graphene in a Strong Magnetic Field,
Rev. Mod. Phys. \textbf{83}, 1193 (2011).

\end{thebibliography}
\end{document}